 \documentclass[11pt]{article}

\usepackage[utf8]{inputenc}
\usepackage[english]{babel}
\usepackage[active]{srcltx}
\usepackage{hyperref}
\usepackage{amsmath,amssymb}
\usepackage[all]{xy}

 1 

\newtheorem{teor}{Theorem}
\newtheorem{prop}{Proposition}
\newtheorem{corol}{Corollary}

\newtheorem{definition}{Definition}

\newtheorem{remark}{Remark}

\def\qedr{\ifvmode\Realemovelastskip\fi
{\unskip\nobreak\hfil\penalty50\hbox{}\nobreak\hfil \hbox{
$\diamond$}\parfillskip=0pt \finalhyphendemerits=0
\par\smallskip}}
\def\ds{\displaystyle}
\newenvironment{proof}{\noindent{\sl Proof:~~~}}{\quad \qed}

\def\beq{\begin{equation}}
\def\eeq{\end{equation}}
\def\bea{\begin{eqnarray}}
\def\eea{\end{eqnarray}}
\def\beann{\begin{eqnarray*}}
\def\eeann{\end{eqnarray*}}
\def\beasn{\begin{sneqnarray}}
\def\eeasn{\end{sneqnarray}}
\def\ben{\begin{enumerate}}
\def\een{\end{enumerate}}
\def\bit{\begin{itemize}}
\def\eit{\end{itemize}}

\def\derpar#1#2{\displaystyle\frac{\partial{#1}}{\partial{#2}}}

\def\vf{{\mathfrak{X}}}
\def\df{{\mit\Omega}}

\def\d{{\rm d}}

\def\Nat{\mathbb{N}}

\def\Real{\mathbb{R}}

\def\Lie{\mathop{\rm L}\nolimits}
\def\inn{\mathop{i}\nolimits}
\def\Cinfty{{\rm C}^\infty}

\def\tabaddress#1{{\small\it\begin{tabular}[t]{c}#1
\\[1.2ex]\end{tabular}}}
\def\qed{\ifvmode\removelastskip\fi
{\unskip\nobreak\hfil\penalty50\hbox{}\nobreak\hfil \hbox{\vrule
height1.2ex width1.2ex}\parfillskip=0pt \finalhyphendemerits=0
\par\smallskip}}

\title{A summary on symmetries and conserved quantities of autonomous Hamiltonian systems
}

\author{{\sc Narciso Rom\'an-Roy\thanks{{\bf e}-{\it mail}:
   narciso.roman@upc.edu  / ORCID: 0000-0003-3663-9861.}}
   \\
   \tabaddress{Departament of Mathematics.
Universidad Polit\'ecnica de Catalu\~na. \\
   Edificio C-3, Campus Norte UPC.
   C/ Jordi Girona 1. 08034 Barcelona. Spain}}

\begin{document}

\maketitle

\pagestyle{myheadings}

\thispagestyle{empty}

\begin{abstract}
A complete geometric classification
of symmetries of autonomous Hamiltonian systems is established; 
explaining how to obtain their associated conserved quantities in all cases.
In particular, first we review well-known results and properties 
about the symmetries of the Hamiltonian and of the symplectic form and then 
some new kinds of non-symplectic symmetries and their conserved quantities are introduced and studied.
\end{abstract}


\vbox{\raggedleft AMS s.\,c.\,(2010): 37J15, 37K05, 53D05, 70H05, 70S05, 70S10.}\null
\markright{\rm N. Rom\'an-Roy,
   \sl Symmetries of Hamiltonian systems.}

\section{Introduction}
\label{intro}

The existence of symmetries of Hamiltonian and Lagrangian systems is related with
the existence of {\sl conserved quantities} (or {\sl constants of motion\/}).
All of them allow us to simplify the integration of dynamical equations,
applying suitable reduction methods \cite{AM-fm,Ar,LM-sgam,MR-99}.
The use of geometrical methods is a powerful tool 
in the study of these topics.
In particular, we are interested in the case of {\sl regular} (i.e., symplectic) Hamiltonian systems.
The most complete way to deal with these problems  
is using the {\sl theory of actions of Lie groups on  (symplectic) manifolds},
and the subsequent theory of reduction  \cite{AM-fm,LM-sgam,MSSV-85,MW-74}
(see also \cite{EMR-99,MRSV-2015} for a extensive list of references that 
cover many aspects of the problem of reduction by symmetries
in a lot of different situations).
Nevertheless, 
the problem of reduction will not be addressed in this dissertation.

As it is well known, the standard procedure to obtain conserved quantities consists in introducing the
so-called {\sl Noether symmetries}, and then
use the {\sl Noether theorem} which is stated both for 
the Lagrangian and the Hamiltonian formalism
in mechanics (and field theories). 
Noether's theorem gives a procedure to associate conservation laws to Noether symmetries
\cite{AM-fm,Ar,LM-sgam,MM-86,MR-99}. 
However, these kinds of symmetries do not exhaust the set of
symmetries. As is known, there are
symmetries which are not of Noether type, but they
also generate conserved quantities
(see, for instance, \cite{CI-83,CMR-2002,Ch-2003,Ch-05,Cr-83,LMR-99,Lu-80,Lu-90,Ra-95,Ra-00}),
and they are sometimes called {\sl hidden symmetries}. 
Different attempts have been made to extend Noether's results or state new theorems in order
to include and obtain the conserved
quantities corresponding to these symmetries, for dynamical systems (for instance, see \cite{BT-2012,Ch-05,Lu-90,Lu-82,SC-81,SC-81b}) and also for field theories \cite{EMR-99b,GPR-2016,RSV-13}.

The aim of this paper is to make a broad summary about the geometric study of symmetries 
of dynamical Hamiltonian systems (autonomous and regular)
in the environment of symplectic mechanics.
In particular, we establish a complete scheme of classification
of all the different kinds of symmetries of Hamiltonian systems, explaining how
to obtain the associated conserved quantities in each case.
We follow the same lines of argument as in the analysis made in  \cite{SC-81} 
for nonautonomous Lagrangian systems,
where the authors obtain conserved quantities for different kinds of symmetries 
that do not leave the {\sl Poincar\'e-Cartan form} invariant.

In particular, in Section \ref{hamsys},
after stating the main concepts about the geometric (symplectic) description
of (autonomous) Hamiltonian systems, we introduce the concept and characterization of
symmetries and conserved quantities and we classify the symmetries in two groups: 
those leaving invariant the geometric structure (the symplectic form),
which are called {\sl geometric symmetries},
and those leaving invariant the dynamics (the Hamiltonian function),
which are called {\sl Hamiltonian symmetries}.
Then, we review 
{\sl Noether symmetries}; that is, those which are both geometrical and Hamiltonian,
and their conserved quantities; stating the {\sl Noether theorem\/}  and its inverse \cite{AM-fm,LM-sgam,MR-99}.
Section \ref{NnN} is devoted to
study {\sl non-Noether symmetries}.
First, non-Hamiltonian symmetries are also reviewed, 
explaining how to obtain their associated conserved quantities,
depending on whether the symmetry is or not geometric too.
The most original part of the paper is in Section \ref{ngs},
where different kinds of non-geometric symmetries are defined,
depending on how the symplectic form transforms under the symmetry.
All of them are studied in detail, showing how to obtain conserved quantities
depending on whether the symmetry is or not Hamiltonian.
Finally, in Section \ref{examples} we present some typical examples
of dynamical systems 
that illustrate some of the cases presented.

All manifolds
are real, paracompact, connected and $C^\infty$. All
maps are $C^\infty$. Sum over crossed repeated indices is understood.


\section{Symplectic mechanics}
\label{hamsys}

 \subsection{Hamiltonian systems. Symmetries. Conserved quantities}

(See, for instance, \cite{AM-fm,Ar,Cr-77,Jo-2016,LM-sgam,MR-99,MW-74}
for more information on the topics in this section).

\begin{definition}
 \label{Hamsyst}
A {\rm (regular)  Hamiltonian system} is a triad
$(M,\omega,{\rm h})$, such that
$(M,\omega)$ is a symplectic manifold; where $M$ represents the {\sl
phase space} of a dynamical system, and
${\rm h} \in\Cinfty(M)$ is the {\rm Hamiltonian function}, which
 gives the dynamical information of the system
(and can be locally or globally defined).
  If $\omega$ is a degenerate form (i.e.; a presymplectic form), then
$(M,\omega,{\rm h})$ is said to be a  {\rm non-regular} (or {\rm singular}\/) {\rm  Hamiltonian system} 
     \end{definition}

Usually $M=T^*Q$, where $Q$ is the {\sl configuration space} of the system.

In this paper, only regular Hamiltonian systems are considered.
In these cases there exists a unique vector field
$X_{\rm h}\in\vf(M)$, which is the {\sl Hamiltonian vector field}
associated with ${\rm h}$:
  \beq
  \label{hameq}
  \inn(X_{\rm h})\omega=\d{\rm h} \ ,
  \eeq
and the dynamical trajectories are the integral curves $\sigma \colon \Real\to M$
of this Hamiltonian vector field $X_{\rm h}\in\vf(M)$.
In a chart of symplectic (Darboux) coordinates $(U;q^i,p_i)$ in $M$ we have that
$\ds X_{\rm h}\mid_U =\derpar{{\rm h}}{p_i}\derpar{}{q^i}-\derpar{{\rm h}}{q^i}\derpar{}{p_i}$,
and the integral curves $\sigma(t)$  of $X_{\rm h}$
are the  solution to the {\sl Hamilton equations}.
If $\dot\sigma\colon\Real\to TM$ denotes the canonical lifting of $\sigma$ to
the tangent bundle $TM$, as 
$\dot\sigma=X_{\rm h}\circ\sigma$,
Hamilton's equations can be written in an intrinsic way
(equivalent to  \eqref{hameq}) as
$$
  \inn(\dot\sigma)(\omega\circ\sigma)=\d{\rm h}\circ\sigma \ .
$$

\begin{definition}
A {\rm symmetry} (or a {\rm dynamical symmetry\/}) of a Hamiltonian system is 
a diffeomorphism $\Phi\colon M \to
M$ such that, if $\sigma$ is a solution to  the Hamilton equations,
then $\Phi\circ\sigma$ is also a solution;
which is equivalent to  $\Phi_*X_{\rm h}=X_{\rm h}$.
If $M=T^*Q$ and $\Phi=T^*\varphi$
(\,$\Phi$ is the canonical lifting of a diffeomorphism  $\varphi\colon Q\to Q$
to the cotangent bundle), then
$\Phi$ is a {\rm natural symmetry}.

 An {\rm infinitesimal symmetry}  (or a {\rm infinitesimal dynamical symmetry\/})
 is a vector field $Y\in{\mathfrak{X}}(M)$ whose local flows are local symmetries.
which is equivalent to $\Lie(Y)X_{\rm h}=[Y,X_{\rm h}]=~0$.
If $M=T^*Q$ and $Y=Z^{C*}$ 
(\,$Y$ is the canonical lifting of $Z\in{\mathfrak{X}} (Q)$
to the cotangent bundle), then $Y$ is a {\rm natural infinitesimal symmetry}.
\end{definition}

\begin{definition}
$f \in \Cinfty (M)$ is a {\rm conserved quantity}
(or a {\rm constant of motion}) if \ $\Lie (X_{\rm h}) f = 0$.
\end{definition}

In particular, the Hamiltonian function ${\rm h}$ is a conserved quantity since
$ \Lie (X_{\rm h}){\rm h} = \inn (X_{\rm h}) \d{\rm h} =\inn^2 (X_{\rm h})\omega= 0$
{\sl (conservation of energy)}.
Furthermore, it is immediate to prove that:

\begin{prop}
if $\Phi\colon M\to M$ is a symmetry and $f\in\Cinfty(M)$ is a
conserved quantity, then $\Phi^*f$ is a conserved quantity.
As a consequence,
if $Y\in\vf(M)$ is an infinitesimal symmetry and $f\in\Cinfty(M)$ is
a conserved quantity, then $\Lie(Y)f$  is a conserved quantity.
\label{generador}
\end{prop}

A symmetry of a Hamiltonian system leaves
the Hamiltonian vector field, $X_{\rm h}$,  invariant. But, as
$X_{\rm h}$ is determined by the geometrical structure
(the symplectic form) and the dynamics
(the Hamiltonian function) through the equation (\ref{hameq}),
some relationship is to be expected between the invariance 
of $X_{\rm h}$ and the invariance of these two elements.
This leads to define:

\begin{definition}
A diffeomorphism $\Phi\colon M\to M$ is a
 {\rm geometric symmetry} of the Hamiltonian system if 
 $\Phi^*\omega=\omega$ (that is, $\Phi$ is a symplectomorphism).

A vector field $Y\in\vf(M)$ is an {\rm infinitesimal geometric symmetry} if
$\Lie(Y)\omega=0$ (that is, it is a local Hamiltonian vector field, $Y\in\vf_{lh}(M)$).

A diffeomorphism $\Phi\colon M\to M$ is a
{\sl Hamiltonian symmetry} if $\Phi^*{\rm h}={\rm h}$.
 
 A vector field $Y\in\vf(M)$ is an {\rm infinitesimal Hamiltonian symmetry} if
 $ \Lie(Y){\rm h}=0$.
\end{definition}

\begin{prop}
\label{refe}
Every (infinitesimal) geometrical and Hamiltonian symmetry is a (infinitesimal) symmetry.
\end{prop}
 {(\sl Proof\/})\
$0=\Phi^*(\inn(X_{\rm h})\omega-\d{\rm h})=\inn(\Phi_*^{-1}X_{\rm
h})\Phi^*\omega-\Phi^*\d{\rm h}= \inn(\Phi_*^{-1}X_{\rm
h})\omega-\d{\rm h}=\inn(X_{\rm h})\omega-\d{\rm h} \\
 \Longrightarrow\ \Phi_*^{-1}X_{\rm h}=X_{\rm h}$.

\quad $\inn([Y,X_{\rm h}])\omega=\Lie(Y)\inn(X_{\rm h})\omega-\inn(X_{\rm
h})\Lie(Y)\omega=\Lie(Y)\d{\rm h}=\d\Lie(Y){\rm h}=0 \ \Longrightarrow\ 
[Y,X_{\rm h}]=0$.
\qed

\subsection{Noether symmetries. Noether's theorem}
\label{one}

(See, for instance, 
\cite{AM-fm,Ar,Jo-2016,LM-sgam,MM-86,MR-99} for more details on these topics).

\begin{definition}
 \label{symH}
Let $(M,\omega,{\rm h})$ be a Hamiltonian system.

A {\rm Noether symmetry}
is a diffeomorphism $\Phi\colon M\to M$ such that:

{\rm (i)} $\Phi^*\omega=\omega$ \quad ; \quad
{\rm (ii)} $\Phi^*{\rm h}={\rm h}$.

\noindent If $M=T^*Q$ and $\Phi=T^*\varphi$, for
a diffeomorphism  $\varphi\colon Q\to Q$, then
$\Phi$ is a {\rm natural Noether symmetry}.

 An {\rm infinitesimal Noether symmetry}
 is a vector field $Y\in\vf(M)$  such that: 

{\rm (i)}  $\Lie(Y)\omega=0$ \quad ; \quad
{\rm (ii)} $\Lie(Y){\rm h}=0$.

\noindent If $M=T^*Q$ and $Y=Z^{C*}$, for $Z\in{\mathfrak{X}} (Q)$, then $Y$ is
a {\rm natural infinitesimal Noether symmetry}.
\end{definition}

 Thus, a (infinitesimal) Noether symmetry is a (infinitesimal) geometric and Hamiltonian symmetry
 and hence it is a symmetry.
From now on we consider only infinitesimal symmetries.

\begin{teor} {\rm (Noether)}:
 Let $Y\in\vf(M)$ be an infinitesimal Noether symmetry.
\begin{enumerate}
\item
The form $\omega_{(0)}\equiv\inn(Y)\omega\in\df^1(M)$ is closed.
Then, for every $p\in M$, there is $U_p\ni p$, there exists
$f_Y\in\Cinfty(U_p)$, unique up to a constant function, such that
$\inn(Y)\omega=\d f_Y$ (on $U_p$).
\item
$f_Y$ is a conserved quantity on $U_p$; that is,
 $\Lie(X_{\rm h})f_Y=0 $.
\end{enumerate}
\label{NTh}
\end{teor}
 {(\sl Proof\/})\ 
1. $\d\inn(Y)\omega=\Lie(Y)\omega-\inn(Y)\d\omega=0$.

\qquad 2. $\Lie(X_{\rm h})f_Y = \inn(X_{\rm h})\d f_Y = \inn(X_{\rm
h})\inn(Y)\omega= -\inn(Y)\inn(X_{\rm h})\omega = -\inn(Y)\d{\rm
h}=-\Lie(Y){\rm h}=0$. \qed

\begin{corol}
The function $f_Y$ is invariant by $Y$.
(No new conserved quantities are generated
by the action of $Y$ on $f_Y$). 
\end{corol}
  {(\sl Proof\/})\
As $f_Y$ is a Hamiltonian function of $Y$, then
 $\Lie(Y)f_Y=\inn(Y)\d f_Y = \inn^2(Y)\omega= 0$.
\qed

 \begin{corol}
If $\omega=\d\theta$, for $\theta\in\df^1(U_p)$, then
there exists $\xi_Y\in\Cinfty(U_p)$
verifying that \ $\Lie(Y)\theta=\d\xi_Y$, on $U_p$;\ and then
$f_Y=\xi_Y-\inn(Y)\theta$\  (up to a constant function).
\end{corol}
   {(\sl Proof\/})\
 In $U_p$ we have that
$0=\Lie(Y)\omega= \Lie(Y)\d\theta=\d\Lie(Y)\theta$.
Then there exists $\xi_Y\in\Cinfty(U_p)$
such that $\Lie(Y)\theta=\d\xi_Y$, on $U_p$,
and the result follows from
$$
\d f_Y=\inn(Y)\omega=\inn(Y)\d\theta=
\Lie(Y)\theta-\d\inn(Y)\theta=\d\xi_Y-\d\inn(Y)\theta \ .
$$
\qed

\begin{teor} {\rm (Inverse Noether)}:
\label{invNo}
For every conserved quantity $f\in\Cinfty(M)$, its Hamiltonian vector field
$Y_f\in\vf_{lh}(M)$ is an infinitesimal Noether symmetry.
\end{teor}
 {(\sl Proof\/})\
 As $Y_f\in\vf_{lh}(M)$, then $\Lie(Y_f)\omega=0$. In addition,
 $$
 \Lie(Y_f){\rm h}=\inn(Y_f)\d{\rm h}=\inn(Y_f)\inn(X_{\rm h})\omega=
  -\inn(X_{\rm h})\inn(Y_f)\omega= -\inn(X_{\rm h})\d f=-\Lie(X_{\rm h})f=0\ .
 $$
\qed

\section{Non-Noether symmetries}
\label{NnN}

\subsection{Non-Hamiltonian symmetries}
\label{two}

Now we study all the symmetries which are not of Noether's type
(that is, symmetries which are not Hamiltonian and/or geometrical), and how they
generate conserved quantities.
First we analyze the (infinitesimal) {\sl non-Hamiltonian symmetries}; that is, such that
$$
[Y,X_{\rm h}]=0 \quad , \quad  \Lie(Y){\rm h}\not=0 \ .
$$

\begin{teor}
\label{teor01}
Let $Y\in{\mathfrak{X}}(M)$ be an infinitesimal non-Hamiltonian symmetry. 
Then $f=\Lie(Y){\rm h}$ is a conserved quantity  (which may be trivial).
  \label{nonHamsym}
  \end{teor}
 {(\sl Proof\/})\quad
$\Lie(X_{\rm h})\Lie(Y){\rm h}=
\Lie([X_{\rm h},Y]){\rm h}+\Lie(Y)\Lie(X_{\rm h}){\rm h}=0$.
\qed

\begin{remark}{\rm
Bearing in mind that
$0 =\inn([Y,X_{\rm h}])\omega=
\Lie(Y)\inn(X_{\rm h})\omega-\inn(X_{\rm h})\Lie(Y)\omega$, we get
\beq
\inn(X_{\rm h})\Lie(Y)\omega=\Lie(Y)\inn(X_{\rm h})\omega=
\Lie(Y)\d{\rm h}=\d\Lie(Y){\rm h}=\d f\ ;
\label{bihamsim}
\eeq
(then $X_{\rm h}$ is said to be a {\sl bi-Hamiltonian vector field} for
  $\tilde\omega=\Lie(Y)\omega$ and $f=\Lie(Y){\rm h}$,
and the dynamical system is called a {\sl bi-Hamiltonian system}.
For a deeper analysis 
of bi-Hamiltonian systems and their symmetries and
conserved quantities see, for instance, 
\cite{Bo-92,BB-02,CI-83,CMR-2002,Ch-2003,CST-2000,Ra-00}).
Taking this into account, if in addition $\Lie(Y)\omega=0$, from (\ref{bihamsim}) we obtain that
 $\d{\rm f}=0$, and then $f$ is locally constant.
Furthermore, in the particular case that $\Lie(Y)\omega=c\,\omega$, $c\in\Real$, we have
 $$
 \inn(X_{\rm h})\Lie(Y)\omega=c\,\inn(X_{\rm h})\omega=c\,\d{\rm h} \ ,
 $$
 and (\ref{bihamsim}) lead to the general result that
$f=c\,{\rm h}$ (up to a constant) is the conserved quantity.}
\label{r1}\end{remark}

As a straightforward  consequence of Proposition \ref{generador} we have:

\begin{teor}
\label{corol3}
If $Y\in{\mathfrak{X}}(M)$ is an infinitesimal symmetry
and $\Lie^N(Y){\rm h}=\overbrace{\Lie(Y)\ldots\Lie(Y)}^{N}{\rm h}\not=0$, for $N>1$, then they are conserved quantities (which may be trivial).
\end{teor}


\subsection{Non-geometric symmetries}
\label{ngs}

Next we analyze the (infinitesimal) {\sl non-geometric symmetries};
 that is, such that
$$
[Y,X_{\rm h}]=0 \quad , \quad  \Lie(Y)\omega\not= 0 \ .
$$
Although Theorem \ref{nonHamsym} also applies to
a particular case of this situation (see Remark \ref{r1}), 
there are other possibilities
which we study in the next sections.
Our analysis is based on the methods introduced in \cite{SC-81,SC-81b} for non-Noether symmetries in
the nonautonomous Lagrangian context.

\subsubsection{Higher-order Noether symmetries}

\begin{definition}
 $Y\in{\mathfrak{X}}(M)$ is an
{\rm infinitesimal Noether symmetry of order $N$} if:
\begin{enumerate}
\item
$Y$ is an infinitesimal symmetry.
  \item
There exists $N>1$ such that
  $\Lie^N(Y)\omega= 0$.
 \item
 $\Lie(Y){\rm h}=0$ (that is, $Y$ is a Hamiltonian symmetry).
 \end{enumerate}
\label{CNsymn}
  \end{definition}
  
\begin{remark}{\rm
If condition (3) does not hold ($Y$ is not a Hamiltonian symmetry) then,
as stated in Prop. \ref{generador} and Theor. \ref{corol3},
$\Lie(Y){\rm h}$ and, eventually, $\Lie^m(Y){\rm h}$, for $m>1$, are conserved quantities.
}
\label{r2}\end{remark}

 \begin{teor}
  {\rm (Noether generalized):}
Let $Y\in{\mathfrak{X}}(M)$ be an infinitesimal Noether symmetry of
order $N$. Then:
 \begin{enumerate}
 \item
 The form
  $\Lie^{N-1}(Y)\inn(Y)\omega\in\df^1(M)$ is closed.\\
  Then,  for every $p\in M$, there is a neighborhood $U_p\ni p$
such that
 there exists $f\in\Cinfty(U_p)$, which is unique up to a constant function, satisfying that \
$\Lie^{N-1}(Y)\inn(Y)\omega=\d f$.
\item
The function $f\in\Cinfty(U_p)$
 is a conserved quantity; that is,
 \   $\Lie(X_{\rm h})f=0$ (on $U_P$).
 \end{enumerate}
  \label{Nthgen}
  \end{teor}
   {(\sl Proof\/})\
As $\Lie^N(Y)\omega=0$, we have
$$
\d\Lie^{N-1}(Y)\inn(Y)\omega=\Lie^{N-1}(Y)\d\inn(Y)\omega=
\Lie^N(Y)\omega-\Lie^{N-1}(Y)\inn(Y)\d\omega=0\ .
$$
Furthermore, as $Y\in\vf (M)$ it is a symmetry, then $[Y,X_{\rm h}]=0$ and therefore
 \beann
 \Lie(X_{\rm h})f &=&
 \inn(X_{\rm h})\d f=
 \inn(X_{\rm h})\Lie^{N-1}(Y)\inn(Y)\omega=
 \inn(X_{\rm h})\Lie(Y)\Lie^{N-2}(Y)\inn(Y)\omega
  \\ &=&
 [\Lie(Y)\inn(X_{\rm h}) -\inn([Y,X_{\rm h}])]\Lie^{N-2}(Y)\inn(Y)\omega
=
\Lie(Y)\inn(X_{\rm h})\Lie^{N-2}(Y)\inn(Y)\omega \ ,
 \eeann
 and repeating the reasoning $N-2$ times we arrive at the
 result
 \beann
 \Lie(X_{\rm h})f&=&\Lie^{N-1}(Y)\inn(X_{\rm h})\inn(Y)\omega=
-\Lie^{N-1}(Y)\inn(Y)\inn(X_{\rm h})\omega=
\\ &=&
-\Lie^{N-1}(Y)\inn(Y)\d{\rm h}=
-\Lie^N(Y){\rm h}=0  .
 \eeann
 \qed

\begin{corol}
The function $f$ given in the above theorem is invariant by $Y$.
(No new conserved quantities are generated
by the action of $Y$ on $f$).
\end{corol}
  {(\sl Proof\/})\
  $ \Lie(Y)f=\inn(Y)\d f=\inn(Y)\Lie^{N-1}(Y)\inn(Y)\omega= \Lie^{N-1}(Y)\inn^2(Y)\omega=0$.
\qed

\begin{corol}
If $\omega=\d\theta$, for $\theta\in\df^1(U_p)$, then
there exists $\xi\in\Cinfty(U_p)$
verifying that \ $\Lie^N(Y)\theta=\d\xi$, on $U_p$;\ and then
$f=\xi-\Lie^{N-1}\inn(Y)\theta$\  (up to a constant function).
\end{corol}
   {(\sl Proof\/})\
In fact, we have that, in $U_p$
$$
0=\Lie^N(Y)\omega= \Lie^N(Y)\d\theta=\d\Lie^N(Y)\theta \ ,
$$
then there exists $\xi\in\Cinfty(U_p)$
such that $\Lie^N(Y)\theta=\d\xi$, on $U_p$.
Furthermore,
\beann
\d f&=&\Lie^{N-1}(Y)\inn(Y)\omega=\Lie^{N-1}(Y)\inn(Y)\d\theta=
\Lie^{N-1}(Y)\Lie(Y)\theta-\Lie^{N-1}(Y)\d\inn(Y)\theta
\\ &=&
\Lie^N(Y)\theta-\d\Lie^{N-1}(Y)\inn(Y)\theta=\d\xi-\d\Lie^{N-1}(Y)\inn(Y)\theta
\eeann
and the result follows.
\qed

\subsubsection{Other non-geometric symmetries}
\protect\label{oknns}

If $Y\in\vf(M)$ is not an infinitesimal geometric symmetry and
it is not a higher-order Noether symmetry, then we have that
$\Lie^m(Y)\omega\not=0, \ \forall  m\in\Nat$.
 Then, as the module of $2$-forms in a finite-dimensional manifold is locally finite generated,
 after a finite number of Lie derivations we have that
 the following condition holds (maybe only locally):
\begin{equation}
\Lie^N(Y)\omega=
f_0\,\omega+f_1\Lie(Y)\omega+\ldots+f_{N-1}\Lie^{N-1}(Y)\omega \ ,
\label{efes}
\end{equation}
 being $\omega,\Lie(Y)\omega,\ldots,\Lie^{N-1}(Y)\omega$
 independent forms and  $\{f_0,\ldots,f_{N-1}\}\subset\Cinfty(M)$.
 Therefore:

\begin{teor}
Let $Y\in\vf(M)$ be an  infinitesimal  symmetry  
such that condition (\ref{efes}) holds.
\begin{enumerate}
\item
If $\{f_0,\ldots,f_{N-1}\}\subset\Cinfty(M)$
are not all constant functions, then these non-constant functions
$f_j$ are (non-trivial) local conserved quantities.
\item
If $\{f_0,\ldots,f_{N-1}\}$
are constant functions such that $f_0=0$
(and some of the other are non-vanishing; that is,
$\Lie^N(Y)\omega=C_1\Lie(Y)\omega+\ldots+C_{N-1}\Lie^{N-1}(Y)\omega$),
and $\Lie(Y){\rm h}=0$, (\,$Y$ is an infinitesimal Hamiltonian symmetry),
then:
\begin{enumerate}
\item
 The form
$\gamma\equiv\Lie^{N-1}(Y)\inn(Y)\omega-C_{N-1}\Lie^{N-2}(Y)\inn(Y)\omega-\ldots-C_1\inn(Y)\omega$
is closed. Then, for every $p\in M$, there exist an open
neighbourhood $U_p\ni p$ and a function
 $f\in\Cinfty(U_p)$ (unique up to a constant), such that
$\gamma=\d f$.
\item
$f$ is a local conserved quantity.
\end{enumerate}
\end{enumerate}
\label{teouno}
\end{teor}
{\sl (Proof)}\ 
1.
Remember that $[X_{\rm h},Y]=0$, because $Y$ is an infinitesimal symmetry.
Then, if $\{f_0,\ldots,f_{N-1}\}\subset\Cinfty(M)$ are not all constant functions,
taking Lie derivatives with respect to $X_{\rm h}$ in both sides of the equation (\ref{efes});
for the left-hand side first we observe that
$$
\Lie(X_{\rm h})\Lie(Y)\omega=\Lie([X_{\rm h},Y])\omega+\Lie(Y)\Lie(X_{\rm h})\omega =0 \ ;
$$
then, assuming that $\Lie(X_{\rm h})\Lie^{N-1}(Y)\omega=0$, we obtain
\beq
\Lie(X_{\rm h})\Lie^N(Y)\omega=
\Lie([X_{\rm h},Y])\Lie^{N-1}(Y)\omega+\Lie(Y)\Lie(X_{\rm h})\Lie^{N-1}(Y)\omega=0 \ .
\label{suply}
\eeq
For the right-hand side, bearing in mind (\ref{suply}), we have that
\beann
\Lie(X_{\rm h})(f_{N-1}\Lie^{ N-1}(Y)\omega) &=&
(\Lie(X_{\rm h})f_{N-1})\Lie^{ N-1}(Y)\omega+f_{N-1}\Lie(X_{\rm h})\Lie^{ N-1}(Y)\omega
\\ &=&
(\Lie(X_{\rm h})f_{N-1})\Lie^{ N-1}(Y)\omega  \ .
\eeann
 Then, as $\omega,\Lie(Y)\omega,\ldots,\Lie^{N-1}(Y)\omega$
are independent forms, from (\ref{efes}) and (\ref{suply}) we conclude
$$
\Lie(X_{\rm h})f_0=\ldots =\Lie(X_{\rm h})f_{N-1}=0 \ .
$$

\noindent 2.
Bearing in mind the hypothesis, a direct calculation leads to
\beann
\d\gamma &=&
\Lie^{N-1}(Y)\d\inn(Y)\omega-C_{N-1}\Lie^{N-2}(Y)\d\inn(Y)\omega-\ldots-C_1\d\inn(Y)\omega
\\ &=&
\Lie^N(Y)\omega-C_{N-1}\Lie^{N-1}(Y)\omega-\ldots-C_1\Lie(Y)\omega=0 \ .
\eeann
As $\gamma=\d f$ (locally) and $[Y,X_{\rm h}]=0$, we obtain
 \beann
  \Lie(X_{\rm h})f&=&\inn(X_{\rm h})\d f=\inn(X_{\rm h})\gamma=
\\ &=&
 \inn(X_{\rm h})[\Lie^{N-1}(Y)\inn(Y)\omega-C_{N-1}\Lie^{N-2}(Y)\inn(Y)\omega-\ldots-C_1\inn(Y)\omega]
\\ &=&
[-\inn([Y,X_{\rm h}])+\Lie(Y)\inn(X_{\rm h})]\Lie^{N-2}(Y)\inn(Y)\omega
\\  &&
-C_{N-1}[-\inn([Y,X_{\rm h}])+\Lie(Y)\inn(X_{\rm h})]\Lie^{N-3}(Y)\inn(Y)\omega
\\  &&
-\ldots -C_2[-\inn([Y,X_{\rm h}])+\Lie(Y)\inn(X_{\rm h})]\Lie(Y)\inn(Y)\omega
+ C_1\inn(Y)\inn(X_{\rm h})\omega
\\ &=&
\Lie(Y)\inn(X_{\rm h})[\Lie^{N-2}(Y) -C_{N-1}\Lie^{N-3}(Y)
-\ldots -C_2\Lie(Y)]\inn(Y)\omega+ C_1\Lie(Y){\rm h}
\\ &=&
\Lie(Y)\inn(X_{\rm h})[\Lie^{N-2}(Y)
-C_{N-1}\Lie^{N-3}(Y)
-\ldots -C_2\Lie(Y)]\inn(Y)\omega  \ ,
 \eeann
and repeating the procedure $N-2$ times we arrive to the result
$$
\Lie(X_{\rm h})f=
\Lie^{N-1}(Y)\inn(X_{\rm h})\inn(Y)\omega=
-\Lie^{N-1}(Y)\inn(Y)\inn(X_{\rm h})\omega
=
-\Lie^{N-1}(Y)\inn(Y)\d {\rm h}=
-\Lie^N(Y){\rm h}=0.
$$
\qed

\begin{remark}{\rm
In the remaining cases; that is,
item 2 with $\Lie(Y){\rm h}\not=0$, or when
$$
\Lie^N(Y)\omega=C_0\,\omega+C_1\Lie(Y)\omega+\ldots+C_{N-1}\Lie^{N-1}(Y)\omega 
 \ , {\rm with}\ C_0\not=0 \ ,
$$
we are, in general, in the situation of
Theorems \ref{teor01} and \ref{corol3},
and hence ${\rm h}$ and, eventually, 
$\Lie^N(Y){\rm h}$ (with $N\geq 1$) are conserved quantities.

In the case of item 1, the conserved quantities 
$f_0,\ldots,f_{N-1}$ are not invariant by $Y$ necessarily and
their Lie derivatives could generate new conserved quantities.
In the case 2, no new conserved quantities are generated 
by the action of $Y$ on $f$, since
\beann
\Lie(Y)f&=&\inn(Y)\d f=
\inn(Y)[\Lie^{N-1}(Y)\inn(Y)\omega-C_{N-1}\Lie^{N-2}(Y)\inn(Y)\omega-\ldots-C_1\inn(Y)\omega]
\\ &=&
[\Lie^{N-1}(Y)-C_{N-1}\Lie^{N-2}(Y)-\ldots-C_1]\inn^2(Y)\omega
=0 \ .
\eeann
}\end{remark}

\begin{remark}{\rm
Theorems \ref{Nthgen} and \ref{teouno} give new ways to obtain conserved quantities generated by non-Noether symmetries; 
nevertheless, we are not aware of any examples of  their application.}
\end{remark}

\section{Some examples}
\label{examples}

We illustrate some applications of Theorems \ref{teor01} and \ref{corol3} that yield trivial and nontrivial first integrals respectively in Sections  \ref{example1} and \ref{example2}.
In addition to these ones, other interesting examples of non-Noether symmetries and their associated conserved quantities can be found, for instance, in
\cite{Ch-05,FMP-2001,MN-97,Ra-97,Su-93}
(see also \cite{Pl1,Pl2}, and the references quoted therein, for
another collection of (quantum-mechanical) systems having nontrivial integrals of motion).

\subsection{Example 1: 2-dimensional harmonic oscillator}
\label{example1}

(See also \cite{CMR-2002,LMR-99,SC-81} for this and other similar models).
In this case, $Q=\Real^2$ and $M=T^*Q\simeq \Real^2\times \Real^2$,
with canonical coordinates $(q^1,q^2,p_1,p_2)$ in $T^*Q$,
and the symplectic form reads
\beq
\omega=\d q^1\wedge\d p_1+\d q^2\wedge\d p_2 \ .
\label{symex}
\eeq
Now, the Hamiltonian function is
\beq
{\rm h}=\frac{1}{2}((p_1)^2+(p_2)^2 +(\Omega_1)^2(q^1)^2+(\Omega_2)^2(q_2)^2)\ ,
\label{hfun}
\eeq
where $\Omega_1,\Omega_2$ are constants.
The Hamiltonian vector field is
\beq
 X_{\rm h}=p_1\derpar{}{q^1}+ p_2\derpar{}{q^2}-
(\Omega_1)^2q^1\derpar{}{p_1}-(\Omega_2)^2q^2\derpar{}{p_2} \ .
\label{hvec}
\eeq
This system has two (geometric but non-Hamiltonian) infinitesimal non-Noether symmetries
$$
 Y_1=\frac{\Omega_1}{(\Omega_1)^2(q^1)^2+(p_1) ^2}\left(q^1\derpar{}{q^1}+p_1\derpar{}{p_1}\right) \ , \
Y_2=\frac{\Omega_2}{(\Omega_2)^2(q^2)^2+(p_2) ^2}\left(q^2\derpar{}{q^2}+p_2\derpar{}{p_2}\right) \ ;
$$
in fact, we have that
 $$
\Lie(Y_i){\rm h}=-\Omega_i \quad ,
  \quad \Lie(Y_i)\omega=0 \quad , \quad
   [Y_i,X_{\rm h}]=0\quad ; \quad (i=1,2) \ ;
$$
and the corresponding constants of motion are
  $f_i=\Lie(Y_i){\rm h}=\Omega_i$;
 which, in this case, are constant functions; that is, trivial conserved quantities.
(See also \cite{RD-89} for an analysis of the algebra of symmetries of this model in the case of commensurable frequencies).

\subsection{Example 2: 2-dimensional isotropic harmonic oscillator}
\label{example2}

This is a particular case of the above example, with
$\Omega_1=\Omega_2=\Omega$. Then 
$Q=\Real^2$ and $M=T^*Q\simeq \Real^2\times \Real^2$,
as above,
and the symplectic form is again \eqref{symex}.
The Hamiltonian function is \eqref{hfun}
and the Hamiltonian vector field is \eqref{hvec}
with $\Omega_1=\Omega_2=\Omega$.
For this system, the vector field
$$
 Y=q^2\derpar{}{q^1}+ q^1\derpar{}{q^2}+
p_2\derpar{}{p_1}+p_1\derpar{}{p_2}
$$
verifies that
$$
\begin{array}{ccccc}
[Y,X_{\rm h}]= 0 &,& 
\Lie(Y){\rm h}=2(p_1p_2+\Omega^2q^1q^2) &,&
\\
\Lie(Y)\omega=2(\d q^1\wedge\d p_2+\d q^2\wedge\d p_1)  &,&
\Lie^2(Y)\omega=4(\d q^1\wedge\d p_1+\d q^2\wedge\d p_2)=4\omega &,&
\end{array}
$$
so it is an infinitesimal non-Noether symmetry which is a
non-Hamiltonian and non-geometric symmetry.
Then, according to Theorem \ref{nonHamsym},
a conserved quantity is
$f=p_1p_2+\Omega^2q^1q^2$.
Now we have that $\Lie(Y)f=\Lie^2(Y){\rm h}=4{\rm h}$, and no new conserved quantities arise from $f$.

Nevertheless, it is well known that this dynamical system is 
an example of a {\sl superintegrable system}
\cite{CMR-2002,LMR-99}.
In fact, the Hamiltonian function can be split as
${\rm h}= {\rm h}_1+{\rm h}_2$, where
$\ds {\rm h}_i=\frac{1}{2}\left((p_i)^2+\Omega^2(q^i)^2\right)$ ($i=1,2$),
and  ${\rm h}_1$ and ${\rm h}_2$ are also constants of motion, 
in addition to ${\rm h}$,
since $\Lie(X_{\rm h}){\rm h}_i=0$, for $i=1,2$.
Thus, we have $3=2n-1$ independent conserved quantities
(notice that ${\rm h}_1$, ${\rm h}_2$ and ${\rm h}$ are not independent, but ${\rm h}_1$, ${\rm h}_2$ and $f$ are).

As stated in Theorem \ref{invNo}, there are infinitesimal Noether symmetries 
which originate these new conserved quantities: 
their Hamiltonian vector fields, which are
$$
X_{{\rm h}_i}=p_i\derpar{}{q^i}-\Omega^2q^i\derpar{}{p_i}
\quad ; \quad \mbox{\rm ($i=1,2$)} \ ;
$$
and $X_{\rm h}=X_{{\rm h}_1}+X_{{\rm h}_2}$.
Nevertheless, they can be also associated with other kinds of infinitesimal symmetries.
In fact, the infinitesimal symmetry $Y$ can be split into $Y=Y_1+Y_2$, where
$$
Y_1=q^2\derpar{}{q^1}+p_2\derpar{}{p_1} \quad , \quad
Y_2=q^1\derpar{}{q^2}+p_1\derpar{}{p_2} \ ,
$$
and these vector fields are non-Hamiltonian and non-geometric infinitesimal symmetries. In fact,
$$
\begin{array}{ccccc}
[Y_1,X_{\rm h}]=0 &,& 
[Y_2,X_{\rm h}]=0 &,&  \\
\Lie(Y_1)\omega=\d q^2\wedge\d p_1+\d q^1\wedge\d p_2 &,&
\Lie(Y_2)\omega=\d q^2\wedge\d p_1+\d q^1\wedge\d p_2 &,&  \\
\Lie^2(Y_1)\omega=2\,\d q^2\wedge\d p_2 &,&
\Lie^2(Y_2)\omega=2\,\d q^1\wedge\d p_1 &,& \\
\Lie^3(Y_1)\omega=0 &,&
\Lie^3(Y_2)\omega=0 &,&  \\
\Lie(Y_1){\rm h}=p_1p_2+\Omega^2q^1q^2=f &,&
\Lie(Y_2){\rm h}=p_1p_2+\Omega^2q^1q^2=f &,&  \\
\Lie^2(Y_1){\rm h}=(p_1)^2+\Omega^2(q^1)^2=2{\rm h}_1 &,&
\Lie^2(Y_2){\rm h}=(p_2)^2+\Omega^2(q^2)^2=2{\rm h}_2 &,& \\
\Lie^3(Y_1){\rm h}=0 &,&
\Lie^3(Y_2){\rm h}=0 &.& \\
\end{array}
$$
Therefore, as it is stated in Theorem \ref{corol3},
${\rm h}_1$, ${\rm h}_2$ and $f$ are three independent conserved quantities.

Finally, it is interesting to notice that there are other independent 
non-Noether infinitesimal symmetries having ${\rm h}_1$, ${\rm h}_2$ and $f$
as their associated conserved quantities;
in particular (see \cite{BT-2012}):
$$
Z_1=[(p_2)^2+\Omega^2(q^2)^2]\left(q^2\derpar{}{q^1}+p_2\derpar{}{p_1}\right) \quad , \quad
Z_2=[(p_1)^2+\Omega^2(q^1)^2]\left(q^1\derpar{}{q^2}+p_1\derpar{}{p_2}\right) \ ,
$$
$$
Z_3=[q^1p_2-q^2p_1]\left(p_1\derpar{}{q^1}-p_2\derpar{}{q^2}-
q^1\derpar{}{p_1}+q^2\derpar{}{p_2}\right) \ .
$$

\section{Conclusions and outlook}

A classification of the symmetries
for {\sl (autonomous and regular) Hamiltonian systems} has been done,
obtaining the associated conserved quantities in each case.
In this way, we have reviewed and completed previous results on this topic
(for instance, in \cite{BT-2012,CI-83,Ch-05,LMR-99,Lu-90,Lu-82,Ra-00,SC-81,SC-81b}).

We have reviewed the Noether symmetries 
(which are both geometrical and dynamical)
and the Hamiltonian version of Noether's theorem
(and its converse).
Next, we have considered the non-Noether symmetries.
First, we have analyzed the non-Hamiltonian symmetries
and their conserved quantities;
but the main contribution of the paper is the analysis of
the non-geometric symmetries
(although we are not aware of any instances of application of Theorems
\ref{Nthgen} and \ref{teouno}).
We have seen that there are several types of them,
according to the behaviour of the symplectic structure under the action of the symmetry.
The procedure for obtaining the conserved quantities depends on whether the symmetry is also Hamiltonian or not.
In particular, in some cases, it consist in applying a suitable generalization of the Noether theorem.

A similar study to what we have done here could be done for
{\sl autonomous Lagrangian systems},
although in this case the {\sl  symmetries of the Lagrangian}
must be also considered.
Finally all these results could also be extended to
classical field theories
in order to do a classification of their  symmetries
and the corresponding conservation laws;
completing, in this way, the partial results already obtained in
 \cite{EMR-99b,GPR-2016,RSV-13} for non-Noether symmetries.

\subsection*{Acknowledgments}

{\small

I acknowledge the financial support from the 
Spanish Ministerio de Econom\'{\i}a y Competitividad
project MTM2014--54855--P, 
the Ministerio de Ciencia, Innovaci\'on y Universidades project
PGC2018-098265-B-C33,
and the Secretary of University and Research of the Ministry of Business and Knowledge of
the Catalan Government project 2017--SGR--932.
I also greatly appreciate the comments and suggestions of Prof. Mikhail S. Plyushchay.
Finally, my thanks to the referees for their extensive and valuable comments that have allowed me to significantly improve the final version of the work.

\end{document}